\documentclass[aps,prb,reprint,superscriptaddress,amsmath,amssymb,twocolumn]{revtex4}

\usepackage{graphicx}
\usepackage{dcolumn}
\usepackage{bm}
\usepackage{float}
\usepackage{xcolor}

\begin{document}

\title[Sample title]{Supercell Altermagnets}

\author{Rodrigo Jaeschke-Ubiergo}
\affiliation{Institut f\"{u}r Physik, Johannes Gutenberg-Universit\"{a}t Mainz, Staudingerweg 7, D-55099 Mainz, Germany}
\author{Venkata Krishna Bharadwaj}
\affiliation{Institut f\"{u}r Physik, Johannes Gutenberg-Universit\"{a}t Mainz, Staudingerweg 7, D-55099 Mainz, Germany}

\author{Tomas Jungwirth}
\affiliation{Inst. of Physics Academy of Sciences of the Czech Republic, Cukrovarnick\'{a} 10,  Praha 6, Czech Republic}
\affiliation{School of Physics and Astronomy, University of Nottingham, NG7 2RD, Nottingham, United Kingdom}
\author{Libor Šmejkal}
\affiliation{Institut f\"{u}r Physik, Johannes Gutenberg-Universit\"{a}t Mainz, Staudingerweg 7, D-55099 Mainz, Germany}
\affiliation{Inst. of Physics Academy of Sciences of the Czech Republic, Cukrovarnick\'{a} 10,  Praha 6, Czech Republic}
\author{Jairo Sinova}
\affiliation{Institut f\"{u}r Physik, Johannes Gutenberg-Universit\"{a}t Mainz, Staudingerweg 7, D-55099 Mainz, Germany}
\affiliation{Inst. of Physics Academy of Sciences of the Czech Republic, Cukrovarnick\'{a} 10,  Praha 6, Czech Republic}
\affiliation{Department of Physics, Texas A\&M University, College Station, Texas 77843-4242, USA}

\date{\today}
\begin{abstract}
Altermagnets are compensated magnets with unconventional $d$, $g$, and $i$-wave spin order in reciprocal space.
So far the search for new altermagnetic candidates has been focused on materials in which the magnetic unit cell is identical to the non-magnetic one, i.e. magnetic structures with zero propagation vector. 
Here, we substantially broaden the family of altermagnetic candidates by predicting supercell altermagnets. 
Their magnetic unit cell is constructed by enlarging the nonmagnetic primitive unit cell, resulting in a non-zero propagation vector for the magnetic structure. 
This connection of the magnetic configuration to the ordering of sublattices  gives an extra degree of freedom to supercell altermagnets, which can allow for the control over the order parameter spatial orientation. 
We identify realistic candidates MnSe$_2$ with a $d$-wave order, and RbCoBr$_3$, CsCoCr$_3$, and BaMnO$_3$ with $g$-wave order.  
We demonstrate the reorientation of the order parameter in MnSe$_2$, which has two different magnetic configurations, whose energy difference is only 5 meV, 
opening the possibility of controlling the orientation of the altermagnetic order parameter by external perturbations.
\end{abstract}

\maketitle

Altermagnets are a recently predicted class of magnets with $d$-, $g$-, and $i$-wave spin-order in reciprocal space \cite{Smejkal2021a, Smejkal2022c},
delimited by a classification of the spin symmetries of collinear magnets \cite{Smejkal2021a}. In altermagnets, opposite spin sublattices are connected by a rotation (proper or improper, symmorphic or nonsymmorphic), but cannot be connected by a translation or a center of inversion, making them sharply distinct from conventional antiferromagnets and ferromagnets \cite{Smejkal2021a, Smejkal2022c}. 
The altermagnetic spin symmetries enforce simultaneously magnetic compensation and time-reversal symmetry ($\mathcal{T}$) breaking of the band structure in reciprocal space with alternating spin polarization  \cite{Smejkal2021,Smejkal2022c}. The alternating spin splitting in altermagnets is of nonrelativistic origin, meaning in the limit of no spin-orbit coupling, and can be very large, e.g., in MnTe, CrSb, and RuO$_2$ \cite{Smejkal2021a,Smejkal2020,Ahn2019}. This unconventional spin splitting was recently observed in photoemission studies in MnTe \cite{Krempasky2023,Lee2023,Osumi2023} and RuO$_2$ \cite{Fedchenko2023}. 

Altermagnets have been predicted to host unconventional responses like the crystal anomalous Hall effect without magnetization \cite{Smejkal2020, Smejkal2022AHEReview}, or spin polarized currents with $d$-wave symmetry \cite{Gonzalez-Hernandez2021,Smejkal2021}, that cannot be observed neither in ferromagnets nor antiferromagnets. The anomalous Hall effect was observed in RuO$_2$, MnTe, and Mn$_5$Si$_3$ \cite{Feng2022,Betancourt2021,Reichlova2020, Tschirner2023}, and attributed to the altermagnetic order. Also, the strong spin polarized and spin splitter currents have been observed in RuO$_2$\cite{Bose2022,Bai2021, Karube2022}.

It is well established that altermagnets encompass a diverse range of materials \cite{Smejkal2022c}. However, thus far, all potential candidates exhibit a magnetic unit cell that is identical to the nonmagnetic one. Although this condition is not necessary, when a collinear compensated supercell magnet (where supercell means that the unit cell is enlarged in the magnetic phase) is constructed by doubling the nonmagnetic unit cell, $\mathcal{T}\tau$ emerges as a symmetry. Here $\tau$ represents a non-integer translation of the magnetic unit cell. The collinear magnet is then a conventional antiferromagnet with type IV magnetic space group (MSG), which does not display spin polarization order in the non-relativistic electronic structure. However, not all supercell magnets have type IV MSG. In fact, there are examples of documented magnets with magnetic unit cells larger than the chemical unit cell \cite{Gallego2016a}, exhibiting a type III MSG, which lacks antiunitary translations.

In this article, we identify several magnetic configurations that have been reported by neutron scattering experiments, which are altermagnets with magnetic configurations described by a nonzero propagation vector, meaning that their magnetic unit cells are larger than the nonmagnetic ones. We study the $d$-wave altermagnet MnSe$_2$ with two magnetic configurations \cite{Corliss1958, chattopadhyay1987}, that remarkably show a different orientation of the $d$-wave order parameter for the two different orderings with the same propagation vector $(0,0,1/3)$. We report as well three $g$-wave altermagnets with chemical stoichiometry $AXB_3$. The three candidates have the exact same geometry, but different chemical composition: (i) CsCoCl$_3$ \cite{Melamud1974, Mekata1978}, (ii) RbCoBr$_3$ \cite{minkiewicz1971} and (iii) BaMnO$_3$ \cite{Christensen1972}. Their magnetic structures are described by a non zero propagation vector $(\frac{1}{3},\frac{1}{3},0)$.

\textit{Supercell magnets characterization} | When we refer to a supercell magnet, we are talking about a magnetic material, in which the magnetic unit cell is larger than the chemical unit cell. When describing a magnetic structure, the propagation vectors \cite{Ressouche2014} reflect the translational properties of the magnetic arrangement. A magnetic moment $\boldsymbol{\mu}_{nj}$, where $n$ labels the nonmagnetic unit cells, and $j$ the magnetic species inside each unit cell, can be written as $\boldsymbol{\mu}_{nj} = \sum_{\mathbf{k}} \mathbf{S}_{\mathbf{k}, j} e^{-i\mathbf{k}\cdot \mathbf{R}_n}$. Here, the summation is over the $\mathbf{k}$-vectors of the nonmagnetic Brillouin Zone (BZ). The Fourier coefficients $\mathbf{S}_{\mathbf{k},j}$ are spatial modes of the magnetic structure with propagation vector $\mathbf{k}$. Any supercell magnet has at least one non-zero propagation vector. The coefficients can be written as $\mathbf{S}_{\mathbf{k},j}=S^{\text{(lat)}}(\mathbf{k})\mathbf{S}^{\text{(mag)}}_j(\mathbf{k})$, where $S^{\text{(lat)}}(\mathbf{k})$ is a magnetic lattice structure factor that only has information about the geometrical relation between the magnetic and nonmagnetic unit cells. On the other hand, $\mathbf{S}^{\text{(mag)}}_j(\mathbf{k})$ is a magnetic form factor, that includes the information of the magnetic moments inside the nonmagnetic unit cell \cite{supp-mat-1-2}.

{\emph{Spin symmetry classification of collinear magnetic phases}} | As mentioned above, altermagnets are characterized and classified by their spin symmetries \cite{Smejkal2021a}. These symmetries, in contrast to the conventional magnetic symmetries, involve pairs of generally different operations that act in spin and real space \cite{Litvin1974, Litvin1977}. An element of a spin space group can be written as $[g_{\text{s}}||g_{\text{r}}|\tau]$, where $g_{\text{s}}$ and $g_{\text{r}}$ are point symmetries acting on spin and real spaces respectively, and $\tau$ is a translation. A spin point group (SPG) contains only point symmetries $[g_{\text{s}}||g_{\text{r}}]$. If the SPG contains the real space inversion $[E||\bar{E}]$ ($E$ denotes the identity and $\bar{E}$ the inversion), it is called a spin Laue group (SLG). An arbitrary SPG can always be written as $\mathbf{r}_{\text{S}}\times \mathbf{R}_{\text{S}}$, where $\mathbf{r}_{\text{S}}$ is called the spin-only group, and contains symmetry transformations common to all collinear spin arrangements on crystals \cite{spin-only}, while $\mathbf{R}_{\text{S}}$ is a non-trivial SPG, with elements $[g_{\text{s}}||g_{\text{r}}]$ that are not present in the spin-only group. The nontrivial SLG of a collinear magnet can be used to classify it as one of the three basic magnetic collinear phases \cite{Smejkal2021a}: (i) Ferromagnet; (ii) Antiferromagnet; (iii) Altermagnet. For an altermagnet, it has the structure $\mathbf{R}_{\text{S}}^{\text{III}}=[E||\mathbf{H}] + [C_2||\mathbf{G}-\mathbf{H}]$ where $\mathbf{G}$ is a crystallographic Laue group, $\mathbf{H}$ is a halving subgroup of $\mathbf{G}$, and $C_2$ is a two-fold spin rotation with respect to an axis orthogonal to the collinear spin arrangement. The elements of $\mathbf{H}$ will connect magnetic sites with parallel magnetic moments, while the symmetries in $\mathbf{G}-\mathbf{H}$ will transpose the magnetic sublattices, connecting sites with opposite magnetic moments \cite{supp-mat-1-2}.

  \begin{figure}[ht]
 	\includegraphics[width=\columnwidth]{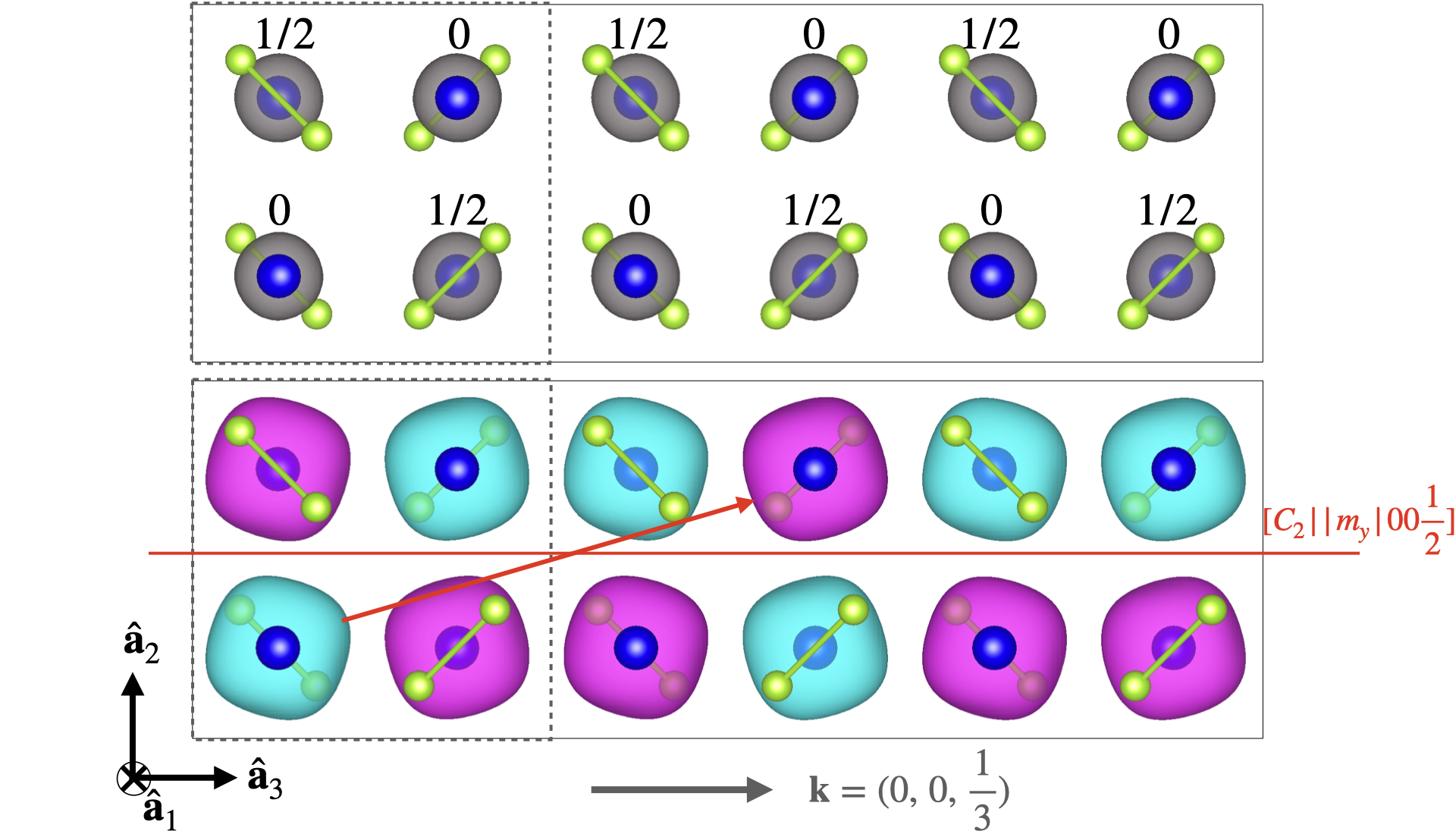}
 	\caption{Magnetic unit cell of MnSe$_2$-I, shown in the plane normal to $(\bar{1}00)$. Blue and green spheres represent Mn and Se atoms, respectively. (top) An isosurface of the charge density is plotted in gray. The fractional coordinates of Mn sites in $\mathbf{a}_1$ ($0$ or $1/2$) are written on top of each atom. (bottom) Isosurfaces of the spin density. Cyan and magenta represent positive and negative values of the spin density, respectively. In both figures, the black dashed line represents the nonmagnetic unit cell. A glide mirror plane (red) connects Mn sites with opposite magnetic moments. \vspace{-0.5 cm}}
 	\label{fig:Fig_1} 
 \end{figure}
 
 \textit{Methods} | The band structure of the presented materials are calculated with density functional theory (DFT), using the Vienna ab initio Simulation Package (VASP) \cite{Kresse1996,Kresse1996a}. 
 The calculations are performed using spin polarized collinear DFT+U (without SOC). The local correlations are treated using the Dudarev method \cite{Dudarev1998}. 
The atomic structures are extracted from MAGNDATA \cite{Gallego2016a,note-methods}.
  For each material we consider several values of the Hubbard parameter $U=0, 1,2,3\; $eV. In all cases, the band structures display spin splitting, with the symmetries imposed by the altermagnetic phase. The specific value of the Hubbard parameter does not affect 
 qualitative properties of the spin splitting, like the spin-momentum locking with $d$- and $g$-wave symmetry, which are present in all calculations, even at $U=0 \text{ eV}$.

\textit{Supercell $d$-wave altermagnet MnSe$_2$} |
Next we focus on the description of two magnetic orders of MnSe$_2$. Both magnetic orders, here termed MnSe$_2$-I \cite{Corliss1958} and MnSe$_2$-II \cite{chattopadhyay1987} are related to the same nonmagnetic phase, which has space group $Pa\bar{3}$ (No. 205), with a cubic unit cell of lattice parameter $a=6.43$ \AA. 
\begin{figure*}[t]
	\centering
	\includegraphics[width=0.8\textwidth]{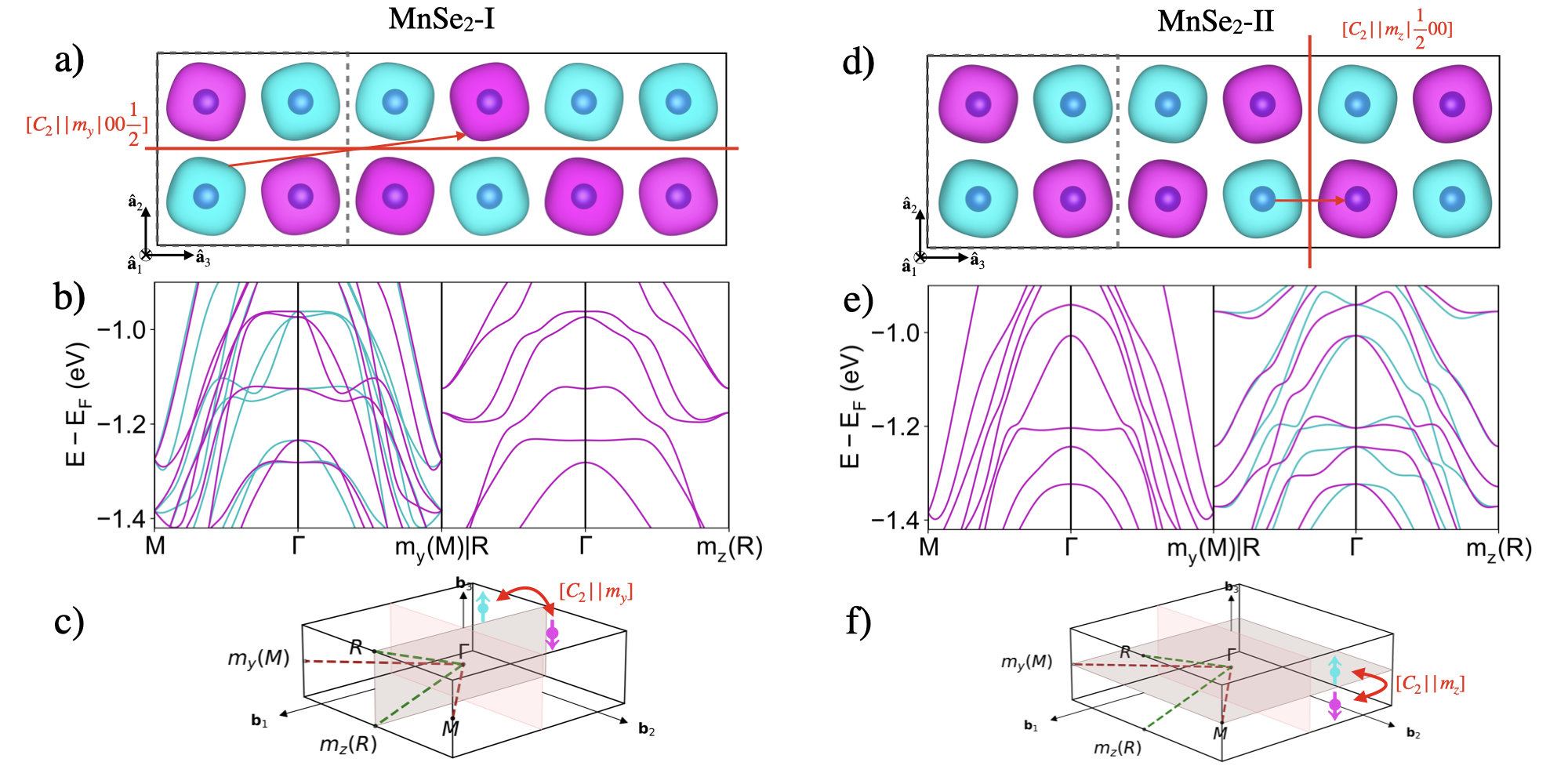}
\caption{Comparison of the magnetic configurations of MnSe$_2$-I (a) and MnSe$_2$-II (d). The magnetic unit cells only differ in the third subcell (from left to right). A glide plane that connects opposite spins is shown in red. (b) and (e) show the band structure along the path $M\Gamma m_y(M)|R\Gamma m_z(R)$. The path is divided into two segments, as can be seen in (c) and (f), denoted by dashed red and green lines. Figures (c) and (f) also show the nodal planes of each magnetic configuration. The different orientation of the $d$-wave causes a spin split path $M\Gamma m_y(M)$ (red) in MnSe$_2$-I, while $R\Gamma m_z(R)$ (green) is spin degenerate, and vice versa for MnSe$_2$-II.}
\label{fig:Fig_2}	
\end{figure*}

Both magnetic structures MnSe$_2$-I and MnSe$_2$-II, are built by stacking three nonmagnetic unit cells along the $c$-axis but, as it will be described later, the distribution of the magnetic moments is different in each case. Figure~\ref{fig:Fig_1} shows the magnetic unit cell of MnSe$_2$-I, where it can be seen how it is built from the nonmagnetic one. The resulting magnetic unit cell is tetragonal, with lattice parameter $c=3a=19.28$ \AA. The electronic charge density (top panel in Fig.~\ref{fig:Fig_1}) has a unit cell which is identical to the nonmagnetic one. Only by looking at the spin density (bottom panel in Fig.~\ref{fig:Fig_1}), it can be seen that the real unit cell is in fact three times larger, and the magnetic configuration can be built from the nonmagnetic unit cell with propagation vectors $(0,0,\pm1/3)$ and $(0,0,0)$ \cite{supp-mat}. Another key aspect that should be noticed in Fig.~\ref{fig:Fig_1} is that the spin density around Mn sites (bottom panel) is anisotropic. This anisotropy illustrates that two sites with opposite spin cannot be connected by a simple antiunitary translation. In this particular case, a glide mirror plane, combined with a two-fold spin rotation (shown in red in Fig.~\ref{fig:Fig_1}) connects two opposite-sign spin densities, protecting the magnetic compensation.

Next we discuss the spin symmetries of  MnSe$_2$-I. The magnetic order does not break inversion symmetry, and therefore the corresponding SPG is already a SLG that can be directly used to identify the collinear magnetic phase. It has the form $^2m ^2m ^1m = [E||2/m] + [C_2|| mmm-2/m] = [E||2/m] + [C_2|| C_{2x}\cdot 2/m]$  \cite{supp-mat}, where $C_{2x}$ is a two-fold rotation with respect to the $[100]$ axis. This group corresponds to a planar $d$-wave altermagnet. The presence of the transposing mirror symmetries $[C_2||m_x]$ and $[C_2||m_y]$,
where $m_{x(y)}$ are real space mirror planes orthogonal to $[100]$ $([010])$, protects two spin degenerate nodal planes in the nonrelativistic band structure. Figure~\ref{fig:Fig_2}a shows the effect of a nonsymmorphic glide plane that acts as a compensation symmetry, connecting spin densities with opposite spin. The resulting $d$-wave nodal planes (normal to $[100]$ and $[010]$ crystal directions) are shown in Fig.~\ref{fig:Fig_2}c.

Next we discuss the second magnetic configuration, MnSe$_2$-II. Here, the magnetic unit cell is also three times larger than the nonmagnetic one, and the magnetic structure has propagation vectors $(0,0,\pm1/3)$ and $(0,0,0)$. However, the distribution of the magnetic moments is  different.  The comparison of configurations I and II can be seen in Fig.~\ref{fig:Fig_2}a and \ref{fig:Fig_2}d. The  difference between the two magnetic configurations is in the two atoms on the far right. This distinction, which at the beginning seems subtle, changes abruptly the spin symmetries of the system. The magnetic order on MnSe$_2$-II breaks inversion symmetry $[E||\bar{E}]$, and the SPG is given by $^2m^2m^12 = [E||2] + [C_2|| mm2-2] = [E||2] + [C_2||m_z\cdot 2 ] $, with $m_z$ a real space mirror plane orthogonal to $[001]$, which is not a SLG. However, in every collinear magnet the symmetry $[C_2 \mathcal{T}||\mathcal{T}]$ (two-fold spin rotation followed by time reversal) is present in the spin-only group $\mathbf{r}_S$, acting effectively as inversion $[E||\bar{E}]$ in the band structure \cite{Smejkal2021a}. This implies that the symmetries of the spin splitting are determined by the SLG, which is given by $\mathbf{R}_S^{\rm III} ={^2m}{^2m}{^12} \times \{[E||E], \;[E||\bar{E}]\}={^2m}{^2m}{^1m}$.
The SLG of MnSe$_2$-II is isomorphic to the SPG of MnSe$_2$-I. However, strictly speaking, the groups are not the same because the axes corresponding to the transposing mirrors $[C_2||m_i]$ are now $[100]$ and $[001]$, as can be seen in Fig.~\ref{fig:Fig_2}c and \ref{fig:Fig_2}f.  This subtle but important change implies a reorientation of the $d$-wave order parameter in reciprocal space.
\begin{figure*}[t]
	\includegraphics[width=0.8\textwidth]{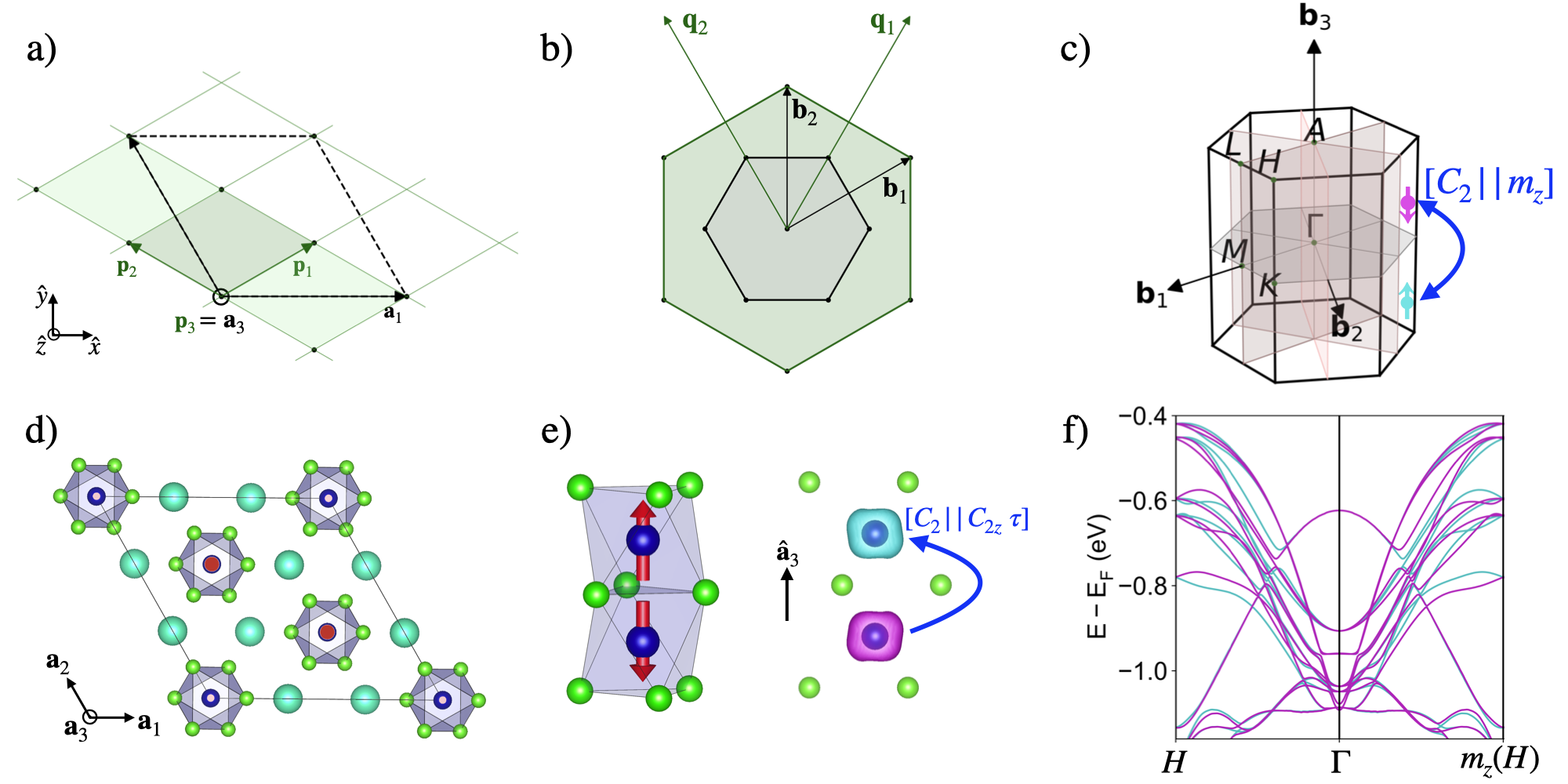}
	\caption{(a) Comparison of magnetic ($\{\mathbf{a}_i\}$) and nonmagnetic ($\{\mathbf{p}_i\}$) unit cells, for the $g$-wave $AXB_3$ supercell altermagnets. (b) BZ (gray) of the magnetic structures $\{\mathbf{b}_i\}$, and BZ (green) of the nonmagnetic phase $\{\mathbf{b}_i\}$. Only $k_z=0$ plane is shown for simplicity. (c) Four nodal planes in the BZ of the $g$-wave altermagnets $AXB_3$, each plane divides domains with opposite spin polarization in the band structure. (d) Crystal structure of the three $g$-wave altermagnets $AXB_3$. $X=$ Mn, Co is the magnetic atom, represented by a blue sphere with a red magnetic moment (pointing out/in the $xy$ plane.) $B$= O, Cl, Br are represented by small green spheres, and $A=$ Ba, Cs, Rb shown as big cyan spheres. (e) $XB_3$ octahedron around two magnetic atoms with opposite magnetic moment. A spin density isosurface around the magnetic atoms is shown in cyan/magenta. (f) The DFT band structure of BaMnO$_3$ ($U=2\text{ eV}$) shows opposite sign spin splitting in the paths $H\Gamma$ and $\Gamma m_z(H)$, since $m_z$ is a transposing symmetry. See Supplementary Material for the band structure of the other two $AXB_3$ materials.}
	\label{fig:Fig_3}
\end{figure*}

To corroborate our symmetry analysis, we perform a spin polarized DFT calculation of the band structure for both configurations along the path $M,\Gamma,m_y(M)| R,\Gamma,m_z(R)$, where $m_y(M)$ and $m_z(R)$ denote the mirrored points with respect to the nodal planes highlighted in Fig.~\ref{fig:Fig_2}. A comparison of the nonrelativistic band structures can be seen in Fig.~\ref{fig:Fig_2}b and \ref{fig:Fig_2}e. We see here that the band structure of MnSe$_2$-I is degenerate along the path $R,\Gamma, m_z(R)$, which is contained in the nodal plane protected by symmetry $[C_2||m_y]$. In the rest of the path, spin splitting is allowed, with a maximum value of 167 meV. On the other hand, the band structure of MnSe$_2$-II is spin degenerate in the path $M,\Gamma,m_y(M)$, because of the symmetry $[C_2|| m_z]$. We see splitting in the rest of the path, with a maximum value of 71 meV. Furthermore, in both cases, the spin split segments show opposite spin polarization at each side of $\Gamma$, consistent with the altermagnetic symmetry. By comparing the total energy (at the same value of the Hubbard parameter $U=2\; \text{eV}$), we see that MnSe$_2$-I is energetically favorable. However, the energy difference is only $5\;\text{meV}$. Such a small difference suggests that it would be possible to tune the ground state by an external perturbation like strain.

\textit{Supercell $g$-wave altermagnets $AXB_3$} |
Next we discuss three $g$-wave altermagnets with the same geometry and spin space group. The materials belong to the family $AXB_3$ with $X$ being a magnetic transition metal:  CsCoCl$_3$ \cite{Melamud1974,Mekata1978}, RbCoBr$_3$ \cite{minkiewicz1971} and BaMnO$_3$ \cite{Christensen1972}. In these materials, the magnetic unit cell is also a supercell of the nonmagnetic one, and it is three times larger. However the geometry of the enlargement is different from MnSe$_2$. Figures~\ref{fig:Fig_3}a and \ref{fig:Fig_3}b show a comparison of both magnetic and non magnetic unit cells, and the respective BZs. From this last scheme, it can be seen that $\mathbf{K}$ and $\mathbf{K'}$ points of the nonmagnetic BZ (vertices of the green larger hexagon) are lattice vectors of the magnetic reciprocal lattice ($\mathbf{b}_i$). This particular enlargement of the nonmagnetic unit cell implies the possible presence of propagation vectors $(0,0,0)$ and $(\pm \frac{1}{3},\pm \frac{1}{3},0)$.

In these materials, the magnetic atoms $X=$ Mn Co, form two triangular lattices stacked along the $c$-axis. The arrangement of the magnetic moments of these atoms is illustrated in Fig.~\ref{fig:Fig_3}d and \ref{fig:Fig_3}e. From a spin symmetry analysis, we recognize the SPG \cite{supp-mat} $\mathbf{R}_S^{III} = [E||\bar{3}m] + [C_2|| 6/mmm - \bar{3}m] = [E||\bar{3}m] + [C_2|| C_{6,(001)}\bar{3}m]$, which corresponds to a $g$-wave altermagnet. The transposing mirrors on the planes normal to $[100]$, $[010]$, $[110]$ and $[001]$, combined with the translation $\tau=(00\frac{1}{2})$ connect (up to a lattice vector)  magnetic atoms that are neighbours in the $c$-axis. The octahedra around the $X$ atoms with opposite magnetic moment cannot be connected by translation or inversion (Fig.~\ref{fig:Fig_3}e), but rather by one of the transposing symmetries of $\mathbf{G}-\mathbf{H}$.

Each of the symmetries $[C_2|| m]$ with $m$ a mirror, impose a spin degenerate nodal plane in the nonrelativistic band structure. These planes connect BZ segments with opposite spin polarization (see Fig.~\ref{fig:Fig_3}c). The band structure of BaMnO$_3$ in Fig.~\ref{fig:Fig_3}f, shows the alternating spin splitting in segments connected by $[C_2||m_z]$.

\textit{Conclusion.}|
We have reported four supercell altermagnets. The magnetic structures of these materials were previously studied by neutron scattering experiments, and collected in \textit{MAGNDATA} \cite{Gallego2016a}. For each material, we have performed a spin symmetry analysis, and verified our results with \textit{ab initio} calculations that show perfect agreement with our prediction of the spin split band structure, and orientation of the spin degenerate nodal planes. For the first candidate, MnSe$_2$, two magnetic configurations were studied. Interestingly, the spin symmetries of each configuration imply a different orientation of the nodal planes. Our DFT calculations show an energy difference of 5 meV per magnetic unit cell. Such a small energy barrier opens the possibility of controlling the $d$-wave order parameter, by making one of the two configurations more energetically favorable. For future research, possible mechanisms to control the transition between the magnetic orders in compensated systems 
are the application of strain, as shown in MnPSe$_3$ \cite{Ni2021}, and alloying, as predicted for example in FeRh and FeRhPd \cite{Komlev2023}. While the application of an electric field has been used to manipulate the magnetic order in multiferroics  \cite{feng2019electric,Spaldin2008b}, such a strategy here would be quite challenging, since it requires the use of a compatible ferroelectric material to interface with MnSe$_2$, to form a multiferroic heterostructure.

Controlling the orientation of the $d$-wave would make possible to switch on and off spin polarized currents \cite{Gonzalez-Hernandez2021}, which in insulators could be carried by magnons \cite{Smejkal2022b}. This idea goes in the same line as the spontaneous anomalous Hall effect in altermagnets \cite{Smejkal2020}, where by reorienting the Néel vector, the relativistic anisotropy axis is changed. In the present case, it is the nonrelativistic $d$-wave order parameter that can be potentially tuned. It is important to highlight that this reorientation would not be possible in a bipartite altermagnet, i.e., an altermagnet with two opposite magnetic moments in the unit cell. The multiple magnetic sublattices provide an extra degree of freedom to couple to the spatial orientation of the $d$-wave. This is possible because the same nonmagnetic phase (parent space group) can be connected with different spin point groups depending on the selection of the magnetic sublattices.

We have also studied three supercell altermagnets with hexagonal unit cells, and isomorphic stoichiometric structures: CsCoCl$_3$, RbCoBr$_3$ and BaMnO$_3$. These three candidates have SPG $^26/^2m^2m^1m$, and hence their band structures contain four spin degenerate nodal planes, with bulk $g$-wave symmetry.

This work is the first report of what we have named supercell altermagnets, materials with a magnetic unit cell larger than the crystallographic one, with spin symmetries that allow for spin splitting of nonrelativistic origin, and spin-momentum locking with $d$-, $g$- or $i$-wave symmetry.
We hope this work encourages the search of new altermagnetic candidates, with nontrivial geometries of their magnetic unit cells.

\textbf{\textit{Acknowledgments}}| We acknowledge fruitful discussions with Dominik Kriegner. Funded by the Deutsche Forschungsgemeinschaft (DFG, German Research Foundation) - TRR 173 – 268565370 (project A03) and TRR 288 – 422213477 (project A09). TJ, LS, JS acknowledge support by Grant Agency of the Czech Republic grant no. 19-28375X. LS acknowledges the Johannes Gutenberg University Grant TopDyn. TJ acknowledges the support of ERC Advanced Grant \#101095925
and the Czech Ministry of Education OP JAK Grant  \#CZ.02.01.01/00/22\_008/0004594.

\nocite{*}
\bibliography{bibliography}

\begin{thebibliography}{40}
\bibitem[{spi()}]{spin-only}
\bibinfo{note}{$\mathbf{r}_{\rm S}$ can take one of the following forms: (i)
  $SO(2)\times \mathbb{Z}_2$ for collinear magnets; (ii) $\mathbb{Z}_2$ for
  noncollinear coplanar magnets; (iii) $1$ for noncoplanar magnets. Here
  $\mathbb{Z}_2=\{[E||E], [C_2\mathcal{T}|| \mathcal{T}] \}$.}
\end{thebibliography}


\renewcommand{\figurename}{Fig. S}
\renewcommand{\theequation}{S\textup{\arabic{equation}}}
\renewcommand \thesection {SUPPLEMENTARY SECTION--\Roman{section}}
\setcounter{section}{0}

\onecolumngrid

\vspace{1.0 cm}

\begin{center}
    \bf \large Supplementary Material: Supercell Altermagnets
\end{center} 

\section{Spin symmetries and altermagnetism}\label{App-A}
The symmetries that properly describe the magnetic part of an electronic Hamiltonian, the  exchange and crystal fields, without considering spin-orbit coupling (SOC) are spin symmetries. Spin symmetries  \cite{Litvin1974} are in some sense more general that the conventional magnetic symmetries (Shubnikov groups), because they allow for the application of different symmetry transformations on spin and real space. An element of a spin space group is denoted as $[g_{\text{s}}||g_\text{r}]$, where $g_{\text{s}}$ is a rotation (proper or improper) that acts on spin space only, and $g_{\text{r}}=\{r|\tau\}$ , (using Seitz notation) acts in real space with a point symmetry $r$ combined with a translation $\tau$. In general, a spin space group can be decomposed as the product of two groups $\mathbf{r}_{\text{S}} \times \boldsymbol{\mathcal{G}}_{\rm S}$. On one hand, $\mathbf{r}_{\text{S}}$ contains symmetries that are common to 
one of the three families of spin arrangements in crystals: (i) $\mathbf{r}_{\rm S}= SO(2)\times \mathbb{Z}_2$ for collinear magnets; 
(ii) $\mathbf{r}_{\rm S}=\mathbb{Z}_2$ for noncollinear coplanar magnets; (iii) $\mathbf{r}_{\rm S}=1$ for noncoplanar magnets. On the other hand, $\boldsymbol{\mathcal{G}}_{\rm S}$ contains combined spin and real space transformations that are not present in $\mathbf{r}_{\rm S}$. The spin-only group of collinear magnets can be written more explicitly as $\mathbf{r}_{\text{S}}= [\mathbf{C}_{\infty}||E] \times \{[E||E], [C_2\mathcal{T}||\mathcal{T}]\}$, with $\mathbf{C}_{\infty}$ the group of all rotations around the axis of collinearity of the spin arrangement, and  $[C_2 \mathcal{T}|| \mathcal{T}]$ denoting the combination of time reversal with a two-fold spin rotation around an axis perpendicular to the spins. An advantage of this treatment is that the nontrivial spin space group $\boldsymbol{\mathcal{G}}_{\rm S}$ contains only unitary symmetries, since the elements that contain the antiunitary time reversal symmetry $[\mathcal{T}||\mathcal{T}]$ are placed in the spin-only group.

We define $\mathbf{T}=\{[E||t_{\mathbf{R}}],\;\; \mathbf{R}=n_1 \mathbf{a}_1 + n_2 \mathbf{a}_2 + n_3 \mathbf{a}_2 | n_1, n_2, n_3 \in \mathbb{Z}\}$ as the group of translations in a given Bravais lattice with lattice vectors $\{\mathbf{a}_i\}$. We can factorize the non trivial spin space group $\boldsymbol{\mathcal{G}}_{\rm S}$ with respect to the invariant subgroup of translations $\mathbf{T}$ as follows:

\begin{equation}
	\boldsymbol{\mathcal{G}}_{\rm S} = [s_1||g_1]  \mathbf{T} + [s_2||g_2] \mathbf{T}+ \; ...\; + [s_n||g_n] \mathbf{T}
	\label{spin_space_factor}
\end{equation}

The coset representives $[s_i||g_i]$ are formed by a  spin rotation  $s_i$ that for collinear magnets takes values in $\{ E, C_2\}$ and a crystallographic symmetry $g_i=\{r_i| \tau_i\}$ that can be generally formed by a point symmetry $r_i$ and a non integer translation $\tau_i$. If we take only the rotational part of the crystallographic symmetry (omitting the translation $\tau_i$) the elements $[s_i||r_i]$ will form the nontrivial spin point group of the crystal $\mathbf{R}_{\text{S}}$ \cite{Litvin1977}. In collinear magnets, the nontrivial spin point group allows us to distinguish between magnetic phases. In particular, altermagnets are characterized by the following structure of the spin Laue group  \cite{Smejkal2021a}:

\begin{equation}
	\mathbf{R}_{\text{S}}^{\text{III}} = [E||\mathbf{H}] + [C_2||\mathbf{G}-\mathbf{H}] ,
	\label{eq:Rs_III}
	\end{equation}
where $\mathbf{G}$ is a crystallographic point group, and $\mathbf{H}$ a halving subgorup of $\mathbf{G}$. Here, the elements of $\mathbf{H}$ will map magnetic sites into sites with the same magnetic moment. While the complementary set $\mathbf{G}-\mathbf{H}$ will contain symmetries that connect sites with opposite magnetic moments, protecting then the compensation between the magnetic sublattices. The symmetries connecting the same spin sublattice (elements of $\mathbf{H}$) will be called non-transposing symmetries, while the elements of $\mathbf{G}-\mathbf{H}$, connecting sublattices with opposite spin, will be called transposing symmetries. The spin point groups $\mathbf{R}_{\text{S}}^{\text{III}}$ of the altermagnetic class allows for spin split band band structure, even in absence of SOC. However, as the compensation of the magnetic moments is protected by symmetries of the form $[C_2|| g]$,with $g$ a rotation or rotoinversion, this will imply the relation $E(\mathbf{k}, \sigma) = E(g\mathbf{k}, -\sigma)$ where $E(\mathbf{k},\sigma)$ is an energy  eigenvalue at wavevector $\mathbf{k}$ and spin $\sigma$. For those wave vectors that are invariant under the point symmetry $g$, satisfying $\mathbf{k}=g\mathbf{k}$ (up to a reciprocal lattice vector), the band structure will be spin degenerate. Therefore, the transposing symmetries will define a set of nodal surfaces, and moreover, the spin polarization of the  isoenergy surfaces will have well defined $d$-,$g$- or $i$-wave symmetry. On the present work, we describe supercell altermagnets with $d$-wave and $g$-wave symmetry.

\section{Propagation vector in supercell magnets}\label{App-B}
One of the challenges of analyzing magnetic systems is that sometimes the periodicity of the magnetic structure differs from the intrinsic periodicity of the nonmagnetic crystal. To overcome this difficulty, and to connect a given magnetic texture which the crystal in the nonmagnetic phase, the concept of propagation vector is commonly introduced \cite{Ressouche2014}. Let us assume a magnetic crystal, with well localized magnetic moments around some atoms $	\boldsymbol{\mu}_{nj}$, where $n$ labels the unit cell of the nonmagnetic phase, and $j$ the runs over the magnetic species inside that nonmagnetic unit cell. We are dealing with a magnetic crystal, and then the distributions $	\boldsymbol{\mu}_{nj}$ have some periodicity in space. We can use this to expand them in a Fourier series:

\begin{equation}
	\boldsymbol{\mu}_{nj} = \sum_{\mathbf{k}} \mathbf{S}_{\mathbf{k}, j} e^{-i\mathbf{k}\cdot \mathbf{R}_n},
	\label{eq:prop_vec_mu_S}
\end{equation}
where the wave vector $\mathbf{k}$ runs over the first BZ of the primitive nonmagnetic unit cell. The Fourier coefficients $\mathbf{S}_{\mathbf{k}, j}$ represent spatial modes of the magnetic texture at the ion $j$, with propagation vector $\mathbf{k}$. The propagation vectors are giving us information about the translational properties of the magnetic structure, in relation to the nonmagnetic phase. We will label now the magnetic unit cells as $\mathbf{X}_m$. The magnetic unit cell is equal or bigger than the non magnetic one. It is clear that for a commensurate magnetic structure every nonmagnetic unit cell vector can be written uniquely as $\mathbf{R}_n = \mathbf{X}_m + \mathbf{y}_{\alpha}$ where the vector $\mathbf{y}_{\alpha}$ tells us the position of the nonmagnetic unit cell inside the magnetic one. The index $\alpha$ will run over $1,2,...M$, and the volume of the nonmagnetic unit cell will be contained $M$ times inside the magnetic unit cell. Each label $n$ is well identified with a pair $(m, \alpha)$ and  $\boldsymbol{\mu}_{nj}= \boldsymbol{\mu}_{(m\alpha)j}=\boldsymbol{\mu}_{\alpha j}$, because of the periodicity of the magnetic structure. Inverting the Fourier series of \eqref{eq:prop_vec_mu_S} we get:

\begin{equation}
	\mathbf{S}_{\mathbf{k}, j} = \sum_{\mathbf{R}_n} 	\boldsymbol{\mu}_{nj}  e^{i\mathbf{k}\cdot \mathbf{R}_n} =S^{\text{(lat)}}(\mathbf{k})\mathbf{S}^{\text{(mag)}}_j(\mathbf{k})
	\label{eq:prop_vec_S_mu} \text{ .}
\end{equation}
With $S^{\text{(lat)}}(\mathbf{k})=\sum_{\mathbf{X_n}}e^{i\mathbf{k}\cdot \mathbf{X_m}}$ representing a magnetic lattice structure factor, which only depends on the geometrical relation between the magnetic and nonmagnetic unit cells, and $\mathbf{S}^{\text{(mag)}}_j(\mathbf{k})=\sum_{\mathbf{y}_{\alpha}} \boldsymbol{\mu}_{\alpha j} e^{i\mathbf{k}\cdot \mathbf{y}_{\alpha}}$ a magnetic form factor, that depends on the  magnetic moments inside the unit cell.
The set of propagation vectors $\mathbf{k}$ of a given magnetic structure will be determined by those wave vectors in which neither of the terms in the last expression vanish. The magnetic cell form factor $S^{\text{(lat)}}(\mathbf{k})$ will vanish unless $\mathbf{k}$ is a reciprocal lattice vector of the magnetic lattice. This give us a simple rule to see which propagation vectors can be allowed only depending on the geometrical relation between the magnetic and non magnetic unit cells. However, given a particular magnetic configuration, the spin form factor $\mathbf{S}^{\text{(mag)}}_j(\mathbf{k})$ could vanish on those $\mathbf{k}$ vectors, further constraining the set of propagation vectors. For example, a zero propagation vector will be absent when $\sum_{\alpha}\boldsymbol{\mu}_{\alpha j} = 0$ for all magnetic atoms $j$. This would be the case of conventional antiferromagnet with $\mathcal{T}\tau$ symmetry (time reversal combined with translation) in which the nonmagnetic unit cell is doubled, and the magnetic moments on each subcell are antiparallel.

\section{Symmetry generators and details on the crystal structures}\label{App-C}

\textit{$d$-wave supercell altermagnet MnSe}.| The nonmagnetic unit cell is cubic, and the magnetic one is built by stacking three unit cells in the $c$-axis. By looking at the magnetic lattice structure factor $S^{\text{(lat)}}(\mathbf{k})$, it follows that those $\mathbf{k}$ vectors of the nonmagnetic BZ which correspond to reciprocal lattice vectors of the magnetic BZ are $(0,0,0), (0,0,\frac{1}{3})$ and $(0,0,\frac{\bar{1}}{3})$. In this particular structure, the magnetic form factor $\mathbf{S}^{\text{(mag)}}_j(\mathbf{k})$ is finite in all three. In particular, it is simple to see why the zero propagation vector is present, since the Mn atoms located at $(0,0,0), (0,0,\frac{1}{3}), (0,0,\frac{2}{3})$ belong to the same index $j$ in the nonmagnetic unit cell, and have spin orientation $\uparrow, \downarrow$ and $\downarrow$ respectively, not being able to cancel the zero vector mode. Tables \ref{table:spin_space_MnSe2} and \ref{table:spin_space_MnSe2_II} show the generators of the spin space groups of MnSe$_2$-I and MnSe$_2$-II respectively.

\textit{MnSe$_2$-I}.| There are in total 12 Mn atoms in the magnetic unit cell, and by looking at the spin-crystal orbits induced by the spin space group, we recognize that there are two magnetic lattices which are not connected by any symmetry. Each of these lattices, Mn$^{(1)}$ with 4 sites, and Mn$^{(2)}$ with 8 sites, are independently compensated. This means that lattice Mn$^{(1)}$ can be further decomposed into sublattices Mn$_{A}^{(1)}$ and Mn$_{B}^{(1)}$, where sites labeled by $A(B)$ contain atoms with magnetic moment $\uparrow(\downarrow)$. Analogously, the second lattice Mn$^{(2)}$, can be as well subdivided into Mn$_{A}^{(2)}$ and Mn$_{B}^{(2)}$. The coordinates of the magnetic sublattices can be seen in Table \ref{table:Magnetic_lattice_MnSe2}.
It is important to highlight that those symmetries associated with the invariant halving subgroup $\mathbf{H}$, will map the sublattices into themselves, while the rest of the symmetries, will connect Mn$^{(1)}_A$ $\leftrightarrow$ Mn$^{(1)}_B$ and  Mn$^{(2)}_A$ $\leftrightarrow$ Mn$^{(2)}_B$.

\textit{MnSe$_2$-II}.| The symmetry of the spin arrangement II is lower than the symmetry of I, and the decomposition of the magnetic sites into orbits of the spin space group leads to not two, but three symmetry independent magnetic sublattices Mn$^{(1)}$, Mn$^{(2)}$, Mn$^{(3)}$ (see Table \ref{table:Magnetic_lattice_MnSe2-II}). Each contains four atoms, two with spin $\uparrow$ (A), and the other two with spin $\downarrow$ (B). The magnetic sublattices are independently compensated by the transposing symmetries (see for example glide plane $[C_2||m_z|\frac{1}{2}00]$ highlighted in red on Fig. 2d in the main text).

While the compensation of the magnetic moments is protected independently inside each sublattice $(1)$ and $(2)$  (for $MnSe_2$-I) and $(1)$, $(2)$, $(3)$ (for  $MnSe_2$-II), there are no symmetries forcing magnetic moments of different sublattices to have the same magnitude. However, both the experimental evidence and our DFT calculations suggest that all magnetic moments in this material have the same magnitude. This should not be surprising, since even though the sublattices are not connected by any symmetry, they belong to the same Wyckoff position in the non magnetic system, and hence all Mn sites have identical chemical environment and should experience the same magnitude of exchange interactions, leading to equal magnitude moments.

\begin{figure*}
	\centering
	\includegraphics[width=0.95\textwidth]{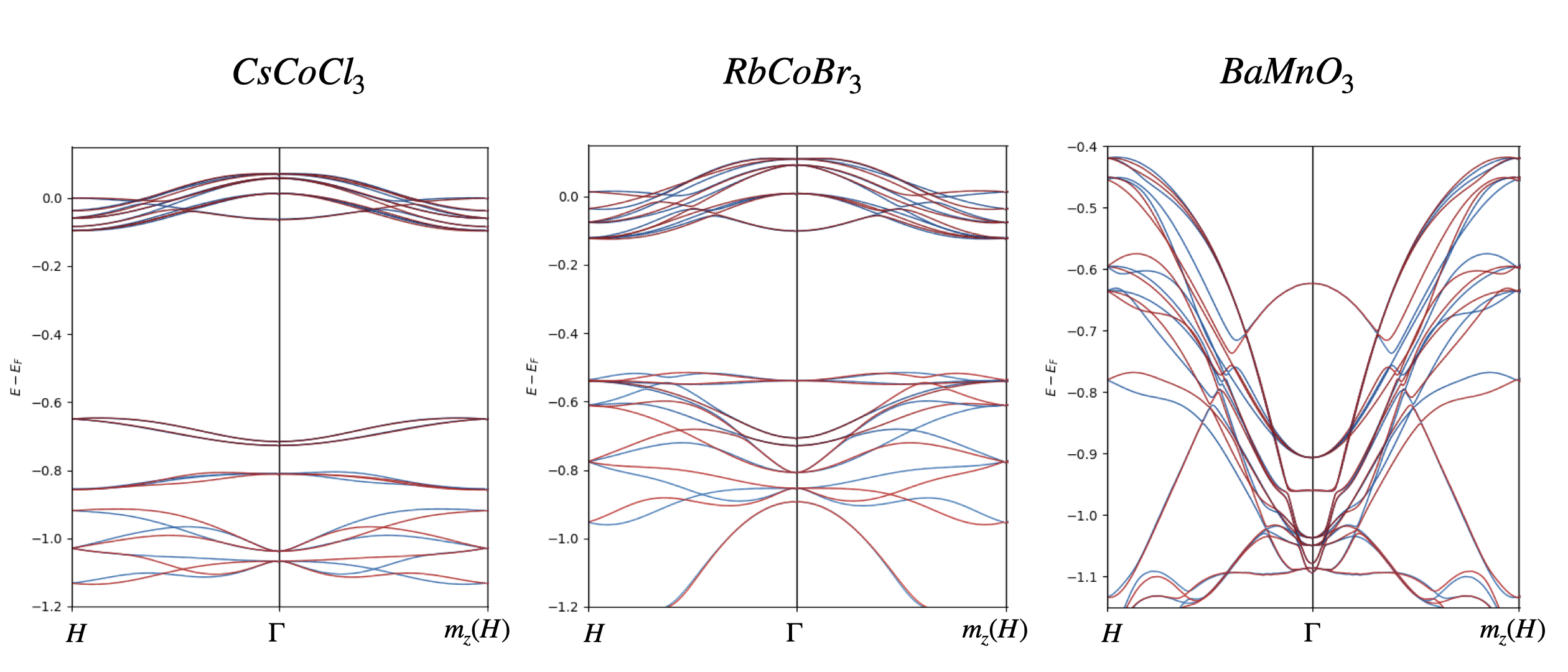}
\caption{Band structures of the three $g$-wave $AXB_3$ supercell altermagnets. Colors red and blue represent opposite spin polarization. Hubbard parameter used was $U=2$ eV.}
\label{fig:Fig_appendix_1}	
\end{figure*}

\textit{$g$-wave altermagnets $AXB_3$}.|
The symmetry generators can be seen in Table \ref{table:spin_space_gwave_generators}. The spin space group allows us to classify the magnetic lattices according to different orbits of the atomic sites. Two types of symmetry independent magnetic atoms $X^{(1)}$ and $X^{(2)}$ are recognized. The sublattice $X^{(1)}$  contains two sites with opposite magnetic moments, located at the boundary of the magnetic unit cell of Fig. 3d in the main text. On the other hand, the sublattice $X^{(2)}$ contains four magnetic atoms. Two of them in the top layer have magnetic moments along $(001)$, while the two atoms in the bottom layer have magnetic moments along $(00\bar{1})$. The coordinates of the magnetic sublattices can be seen in Table \ref{table:magnetic_sublattices_AXB3}. Each of the magnetic sublattices are independently compensated. The calculated band structures for CsCoCl$_3$, RbCoBr$_3$ and BaMnO$_3$ are shown in Fig. \ref{fig:Fig_appendix_1}.

Next we explain why all the propagation vectors mentioned in the main text (including $\mathbf{k}=(0,0,0)$) are necessary to describe the magnetic arrangement. By inspection of the structure factors in \eqref{eq:prop_vec_mu_S}, $S^{\text{(lat)}}(\mathbf{k})$ is nonzero for $\mathbf{K}=(\frac{1}{3},\frac{1}{3},0)$ and $(0,0,0)$. Let us now look closely to the magnetic form factor $\mathbf{S}^{\text{(mag)}}_j(\mathbf{k})$. The index $j$ will take values $1,2$ for the top and bottom layer respectively (layers stacked along c-axis). We will consider three nonmagnetic unit cells that cover the full magnetic unit cell, $\mathbf{R}_0=0$, $\mathbf{R}_1=-\mathbf{p}_2$ and $\mathbf{R}_2=\mathbf{p}_2$, as shown in Fig. 3a. of the main text. Furthermore, as the structure is collinear, we are only interested in the $c$-component of the magnetic moments. We obtain the relations

\begin{equation}
	\mu_{0j} = 2 S_{\mathbf{K}j} + S_{0j}\text{ ,}
	\end{equation}

\begin{equation}
	\mu_{1j} = \mu_{2j} =  - S_{\mathbf{K}j} + S_{0j}\text{ , }
\end{equation}
where $S_{\mathbf{K}j}$ is real given the restriction $\mu_{1j}=\mu_{2j}$. If the zero propagation vector was absent, the structure would still be an altermagnet, but there would be a relation between the magnetic moments of the sublattices $X^{(1)}$ and $X^{(2)}$ constraining them to satisfy $\mu_{0j}=-2\mu_{1j}$ and $\mu_{0j}=-2\mu_{2j}$. In the models described here the absolute value of this magnetic moments is equal or nearly equal, and this makes necessary the zero mode $S_{0j}$. The reasoning behind the magnetic moments on symmetry independent sublattices is analogous to the MnSe$_2$ case. The chemical enviroment (Wyckoff position in the space group describing the nonmagnetic phase) is identical for all magnetic atoms, so even though small differences are allowed by symmetry, they are not observed.

\renewcommand{\arraystretch}{1.5}
\begin{table}[h]
	\centering
\begin{tabular}{|c|c|}
	\hline
	Spin space symmetry & Real space symmetry \\ \hline
	$E$ & $\{E| 000\}$ \\ \hline
	$E$ &  $\{\bar{E}| 000\}$  \\ \hline
	$E$ &  $\{C_{2z}| \frac{1}{2}0\frac{1}{2}\}$  \\ \hline
	$E$ &  $\{m_{z}| \frac{1}{2}0\frac{1}{2}\}$  \\ \hline
	$C_2$ &  $\{C_{2x}| \frac{1}{2}\frac{1}{2}0\}$  \\ \hline
	$C_2$ &  $\{C_{2y}| 0\frac{1}{2}\frac{1}{2}\}$  \\ \hline
	$C_2$ &  $\{m_{x}| \frac{1}{2}\frac{1}{2}0\}$  \\ \hline
	$C_2$ &  $\{m_{y}| 0\frac{1}{2}\frac{1}{2}\}$  \\ \hline
\end{tabular}
\caption{Symmetry elements of the spin space group of MnSe$_2$-I, each row represents one coset representative $[s_i||g_i]$ according to eq. \eqref{spin_space_factor} . Taking only the real space point symmetries we form the group $\mathbf{G}=mmm$, the first four rows form the invariant halving subgroup $\mathbf{H}=2/m$. The cartesian setting $\{x,y,z\}$ used to represent the real space point symmetries, is aligned with the lattice vectors $\mathbf{a}_1$, $\mathbf{a}_2$, $\mathbf{a}_3$. The translations are written as fractions of the lattice vectors.}
\label{table:spin_space_MnSe2}
\end{table}

\begin{table}[h]
	\centering
	\begin{tabular}{|c|c|}
		\hline
		Spin space symmetry & Real space symmetry \\ \hline
		$E$ & $\{E| 000\}$ \\ \hline
		$E$ &  $\{C_{2y}| 0\frac{1}{2}\frac{1}{6}\}$  \\ \hline
		$C_2$ &  $\{m_{x}| \frac{1}{2}\frac{1}{2}0\}$  \\ \hline
		$C_2$ &  $\{m_{z}| \frac{1}{2}0\frac{1}{6}\}$  \\ \hline
	\end{tabular}
	\caption{Symmetry elements of the spin space group of MnSe$_2$-II, each row represents one coset representative $[s_i||g_i]$ according to eq. \eqref{spin_space_factor} . Taking only the real space point symmetries we form the group $\mathbf{G}=mm2$, the first four rows form the invariant halving subgroup $\mathbf{H}=2$. The cartesian setting $\{x,y,z\}$ used to represent the real space point symmetries, is aligned with the lattice vectors $\mathbf{a}_1$, $\mathbf{a}_2$, $\mathbf{a}_3$. The translations are written as fractions of the lattice vectors.}
	\label{table:spin_space_MnSe2_II}
\end{table}

\begin{table}[p]
	\centering
	\begin{tabular}{|c|c|c|}
		\hline
		Magnetic sublattice & Spin orientation & Coordinates\\ \hline
		Mn$^{(1)}_A$&$\uparrow$ &$(0,0,0)$, $(\frac{1}{2},0,\frac{1}{2})$ \\ \hline
		Mn$^{(1)}_B$&$\downarrow$& $(\frac{1}{2},\frac{1}{2},0)$, $(\frac{1}{2},\frac{1}{2},\frac{1}{2})$\\ \hline
		Mn$^{(2)}_A$&$\uparrow$& $(0,\frac{1}{2},\frac{1}{6})$, $(\frac{1}{2},\frac{1}{2},\frac{1}{3})$, $(\frac{1}{2},\frac{1}{2},\frac{2}{3})$, $(0,\frac{1}{2},\frac{5}{6})$ \\ \hline
		Mn$^{(2)}_B$&$\downarrow$& $(\frac{1}{2},0,\frac{1}{6})$, $(0,0,\frac{1}{3})$, $(0,0,\frac{2}{3})$, $(\frac{1}{2},0,\frac{5}{6})$\\ \hline
	\end{tabular}
	\caption{Magnetic sublattices in MnSe$_2$-I. Coordinates are given in terms of the lattices vectors of the magnetic lattice $\mathbf{a}_1=a\mathbf{\hat{x}}$,  $\mathbf{a}_2=a\mathbf{\hat{y}}$,\ and  $\mathbf{a}_3=c\mathbf{\hat{z}}$. The four sublattices are obtained by calculating the orbit of the magnetic sites under the invariant subgroup with point symmetries in $\mathbf{H}=2/m$. If the orbits are calculated on the full spin space group, we obtain two sublattices $\text{Mn}^{(1)}$ and $\text{Mn}^{(2)}$ which are independently compensated.}
	\label{table:Magnetic_lattice_MnSe2}
\end{table}

\begin{table}[p]
	\centering
	\begin{tabular}{|c|c|c|}
		\hline
		Magnetic sublattice & Spin orientation & Coordinates\\ \hline
		Mn$^{(1)}_A$&$\uparrow$ &$(0,0,0)$, $(0,\frac{1}{2},\frac{1}{6})$ \\ \hline
		Mn$^{(1)}_B$&$\downarrow$& $(\frac{1}{2},\frac{1}{2},0)$, $(\frac{1}{2},0,\frac{1}{6})$\\ \hline
		Mn$^{(2)}_A$&$\uparrow$& $(\frac{1}{2},0,\frac{1}{2})$, $(\frac{1}{2},\frac{1}{2},\frac{2}{3})$ \\ \hline
		Mn$^{(2)}_B$&$\downarrow$& $(0,\frac{1}{2},\frac{1}{2})$, $(0,0,\frac{2}{3})$\\ \hline
		
		Mn$^{(3)}_A$&$\uparrow$& $(\frac{1}{2},\frac{1}{2},\frac{1}{3})$, $(\frac{1}{2},0,\frac{5}{6})$\\ \hline
		Mn$^{(3)}_B$&$\downarrow$& $(0,0,\frac{1}{3})$, $(0,\frac{1}{2},\frac{5}{6})$ \\  \hline
		
	\end{tabular}
	\caption{Magnetic sublattices in MnSe$_2$-II. Coordinates are given in terms of the lattices vectors of the magnetic lattice $\mathbf{a}_1=a\mathbf{\hat{x}}$,  $\mathbf{a}_2=a\mathbf{\hat{y}}$,\ and  $\mathbf{a}_3=c\mathbf{\hat{z}}$. The six sublattices are obtained by calculating the orbit of the magnetic sites under the invariant subgroup with point symmetries in $\mathbf{H}=2$.}
	\label{table:Magnetic_lattice_MnSe2-II}
\end{table}

 \renewcommand{\arraystretch}{1.5}
\begin{table}
	\centering
	\begin{tabular}{|c|c|c|}
		\hline
		Magnetic sublattice &  Spin orientation & Coordinates\\ \hline
		$X^{(1)}_A$& $\uparrow$ &$(0,0,0)$ \\ \hline
		$X^{(1)}_B$& $\downarrow$ &$(0,0,\frac{1}{2})$ \\ \hline
		$X^{(2)}_A$& $\uparrow$ &$(\frac{1}{3},\frac{2}{3},\frac{1}{2})$, $(\frac{2}{3},\frac{1}{3},\frac{1}{2})$ \\ \hline
		$X^{(2)}_B$& $\downarrow$ &$(\frac{2}{3},\frac{1}{3},0)$, $(\frac{1}{3},\frac{2}{3},0)$ \\ \hline
	\end{tabular}
	\caption{Magnetic sublattices in the $ABX_3$ materials, with $X=\text{Co},\;\text{Mn}$. The coordinates are given in terms of the lattice vectors $\mathbf{a}_1=a\mathbf{\hat{x}}$, $\mathbf{a}_2=-\frac{a}{2} \mathbf{\hat{x}} + \frac{\sqrt{3}a}{2}\mathbf{\hat{y}}$ and $\mathbf{a}_3=c\mathbf{\hat{z}}$, as shown in Fig 3a on the main text.}
	\label{table:magnetic_sublattices_AXB3}
	\end{table}

\renewcommand{\arraystretch}{1.5}
\begin{table}
	\centering
	\begin{tabular}{|c|c|}
		\hline
		Spin space symmetry & Real space symmetry \\ \hline
		$E$ &  $\{\bar{E}| 000\}$  \\ \hline
		$E$ &  $\{C_{2,(1\bar{1}0)}| 000\}$  \\ \hline
		$E$ &  $\{C_{3,(001)}| 000\}$  \\ \hline

		$C_2$ &  $\{C_{6,(001)}| 00\frac{1}{2}\}$  \\ \hline

	\end{tabular}
	\caption{Generators of the spin space group of $AXB_3$ materials, each row represents one coset representative $[s_i||g_i]$ according to eq. \eqref{spin_space_factor}. The spatial part of the first three rows generates the invariant halving subgroup $\mathbf{H}=\bar{3}m$ of equation \eqref{eq:Rs_III}, and by including the transposing generator of the last row $C_{6z}$, we obtain $\mathbf{G}=6/mmm$.}
	\label{table:spin_space_gwave_generators}
\end{table}

\end{document}